\documentclass[11pt]{article}
\usepackage{amsfonts}
\usepackage{latexsym,amsmath}
\usepackage{amssymb,array}
\RequirePackage[dvips]{graphicx}
\usepackage{graphics}
\usepackage{tikz}
\usepackage{adjustbox}
\usepackage{caption}
\usepackage{calc}
\usepackage{graphics}
\usepackage{longtable}
\usepackage{pdflscape}
\usepackage{xcolor}

\usepackage{authblk}
\usepackage[colorlinks=true]{hyperref}
\hypersetup{urlcolor=blue, citecolor=red}
\makeatletter \oddsidemargin 0in \evensidemargin 0in \textwidth
16cm\textheight 20cm 
\begin{document}
\newcommand{\la}{\lambda}
\newcommand{\eq}{\Leftrightarrow}
\newcommand{\mf}{\mathbf}
\newcommand{\ri}{\Rightarrow}
\newtheorem{t1}{Theorem}[section]
\newtheorem{d1}{Definition}[section]
\newtheorem{n1}{Notation}[section]
\newtheorem{c1}{Corollary}[section]
\newtheorem{l1}{Lemma}[section]
\newtheorem{r1}{Remark}[section]
\newtheorem{no1}{Note}[section]
\newtheorem{e1}{Counterexample}[section]
\newtheorem{ex1}{Example}[section]
\newtheorem{p1}{Proposition}[section]
\newtheorem{cn1}{Conclusion}[section]
\newtheorem{il1}{Illustration}[section]
\newtheorem{re1}{Result}[section]
\renewcommand{\theequation}{\thesection.\arabic{equation}}
\pagenumbering{arabic}
\title {Some Results on Comparisons of Random Extremes having Identical and Non-Identical Components}

\author[1]{Bidhan Modok}
\author[2]{Shovan Chowdhury\footnote{Corresponding author e-mail:meetshovan@gmail.com; shovanc@iimk.ac.in}}
\author[1]{Amarjit Kundu}

\affil[1] {Department of Mathematics, Raiganj University, West Bengal, India}
\affil[2] {Quantitative Methods and Operations Management Area, Indian Institute of Management Kozhikode, India}

\maketitle
\begin{abstract}
In this article, we revisit the paper by Kundu \emph{et al.}~\cite{ku}, presenting new results and insights for both identical and non-identical independent random variables. We derive sufficient conditions for preserving the hazard rate and reversed hazard rate orderings between the random maximum and minimum order statistics, respectively. Our results show that these preservation conditions also hold for independent and identically distributed random variables. We also demonstrate that the findings in Kundu \emph{et al.}~\cite{ku} for the non-identical cases do not apply to the identical cases.  

\end{abstract}
\section{Introduction}
\setcounter{equation}{0}
\hspace*{0.2in} Kundu \emph{et al.}~\cite{ku} presented new results on the comparison of minimum and maximum order statistics from a random number of non-identical and independently distributed ($inid$) random variables. In this article, we revisit their work, offering new results and insights for both $inid$ and $iid$ (independent and identically distributed) setups. Suppose, for $i=1,2,\ldots,n$, $X_i$ and $Y_i$ be two sets of $n$ $inid$ (or $iid$) random variables having distribution (survival) functions $F_i\left(x\right)$ ($\bar F_i\left(x\right)$) and $G_i\left(x\right)$ ($\bar G_i\left(x\right)$) respectively. Also let $F_{n:n}\left(x\right)$ and $G_{n:n}\left(x\right)$ ($\bar F_{n:n}\left(x\right)$ and $\bar G_{n:n}\left(x\right)$) be the distribution (survival) functions of $X_{n:n}$ and $Y_{n:n}$ respectively. The respective quantities for $X_{1:n}$ and $Y_{1:n}$ be $F_{1:n}\left(x\right)$ and $G_{1:n}\left(x\right)$ ($\bar F_{1:n}\left(x\right)$ and $\bar G_{1:n}\left(x\right)$). It is to be mentioned here that $X_{1:n}$ and $X_{n:n}$ are defined as $$ X_{1:n}=\min\left\{X_1, X_2,...,X_n\right\} \text{ and } X_{n:n}=\max\left\{X_1, X_2,...,X_n\right\}.$$ 
Again, for the discrete random variable $N$, let $X_{1:N}$ and $Y_{1:N}$ ($X_{N:N}$ and $Y_{N:N}$) be the minimum (maximum) order statistics having distribution and survival functions $F_{1:N}\left(x\right)$, $ G_{1:N}\left(x\right)$ and $\bar F_{1:N}\left(x\right)$, $\bar G_{1:N}\left(x\right)$ ( $F_{N:N}\left(x\right)$, $ G_{N:N}\left(x\right)$ and $\bar F_{N:N}\left(x\right)$, $\bar G_{N:N}\left(x\right)$) respectively.\\ 
\hspace*{0.3in} Kundu \emph{et al.}~\cite{ku}, in Theorems 3.1 and 3.2, demonstrated that under certain sufficient conditions, if a hazard rate ($hr$) ordering exists between \(X_{1:n}\) and \(Y_{1:n}\), the same $hr$ ordering holds between \(X_{1:N}\) and \(Y_{1:N}\). However, the question of whether a reversed hazard rate ($rh$) ordering is preserved between these random variables remains unanswered in their work. Furthermore, in Theorems 3.3 and 3.4 of the same paper, it is shown that under specific conditions, the $rh$ ordering between \(X_{n:n}\) and \(Y_{n:n}\) is preserved between \(X_{N:N}\) and \(Y_{N:N}\). Here, too, the question arises: will the hazard rate ($hr$) ordering between these random variables be preserved?\\
\hspace*{0.2in} Moreover, the results in Kundu \emph{et al.}~\cite{ku} are derived for $inid$ random variables, which may seem obvious for an $iid$ set-up as well. Now, let us consider $X_1, X_2, \ldots, X_n$ and $Y_1, Y_2, \ldots, Y_n$ as $iid$ random variables having survival functions $\bar F(x)$ and $\bar G(x)$ respectively. It is clear that one of the sufficient conditions in Theorem 3.1 of Kundu \emph{et al.}~\cite{ku}, which states that $\frac{\bar F_{1:n}(x)}{\bar G_{1:n}(x)}$ is increasing in $n\in\mathcal{N}^+,$ reduces to the condition that $\left[\frac{\bar F(x)}{\bar G(x)}\right]^n$ is increasing in $n\in\mathcal{N}^+$, where $\mathcal{N}^+$ denotes the set of all positive integers. Furthermore, the assumption in Theorem 3.1 that $X_{1:n}\leq_{hr}Y_{1:n}$ for all $n\in\mathcal{N}^+$ simplifies to $X\leq_{hr}Y$ for $n=1,$ which implies that $X\leq_{st}Y.$ From the definition of $st$ ordering, this also leads to the conclusion that for all $x\geq 0,$ $\bar F(x)\leq \bar G(x)$, which contradicts the earlier condition that $\left[\frac{\bar F(x)}{\bar G(x)}\right]^n$ is increasing in $n\in\mathcal{N}^+$.
This observation is particularly interesting because, although Theorem 3.1 in Kundu \emph{et al.}~\cite{ku} holds for a general set-up such as $inid$ random variables, it cannot be applied to the case of $iid$ random variables. This raises another question: can a similar conclusion be reached for all the theorems in their paper? These gaps motivate us to explore the existence of alternative sets of sufficient conditions that can address these issues and provide further insights on the $iid$ cases.\\
\hspace*{0.2in} In this article, we derive sufficient conditions for the preservation of $hr$ and $rh$ ordering between the maximum and minimum order statistics, respectively. It is demonstrated that these results also apply to $iid$ random variables. Additionally, no sufficient conditions are found for the preservation of $lr$ ordering between the random extremes in the $iid$ set-up. The organization of the paper is as follows. Some definitions and existing results are provided in Section 2. Section 3 provides the main results. Section 4 concludes the paper. 

\section{Preliminaries}
\setcounter{equation}{0}
\hspace*{0.2 in} Let $X$ and $Y$ be two absolutely continuous random variables with respective supports $(l_X,u_X)$ and $(l_Y,u_Y)$, where $u_X$ and $u_Y$ may be positive infinity, and $l_X$ and $l_Y$ may be negative infinity. Then, $X$ is said to be smaller than $Y$ in $i)$ {\it likelihood ratio order} ($X\leq_{lr}Y$), if 
$\frac{f_Y(t)}{f_X(t)}\;\text{is increasing in }\, t \in ((l_X,l_Y)\cup(u_x,u_y)),$ $ii)$ {\it hazard rate order} ($X\leq_{hr}Y$), if $\frac{\bar F_Y(t)}{\bar F_X(t)}\;\text{is increasing in }\, t \in (-\infty,\max(u_x,u_y))$ or equivalently $r_X(t)\geq r_Y(t),$ $ii)$ {\it reversed hazard rate order} ($X\leq_{rh}Y$), if $ \frac{F_Y(t)}{ F_X(t)}\;\text{is increasing in }\, t \in (\min(l_X,l_Y),\infty)$ or equivalently $\overline r_X(t)\leq \overline r_Y(t),$ $iv)$ {\it usual stochastic order} ($X\leq_{st}Y$), if $\bar F_X(t)\leq \bar F_Y(t)$ for all $t \in\mathcal{R} .$
For more details, one can refer to Shaked and Shanthikumar \cite{shak}.\\
\hspace*{0.3in} The next theorem due to Shaked and Shanthikumar \cite{shak} and the lemma are used to prove the key findings of the paper.
\begin{t1}\label{theo11}
 
If $X_1,\ X_2\ldots,\ X_m$ are independent random variables, then  $X_{(k:m-1)}\geq_{rh}X_{(k:m)}$ for $k=1,2,\ldots, m-1.$
\end{t1}

\begin{l1}\label{lemma12}
	For all $0\leq u\leq 1$, $\frac{x}{u^{-x}-1}$ is decreasing in $x\geq 0$.
\end{l1}
\hspace*{0.3in} The notion of Totally Positive of order $2$ ($TP_2$) and Reverse Regular of order $2$ ($RR_2$) are of great importance in various fields of applied probability and statistics (see Karlin \cite{ka}; Karlin and Rinott \cite{ka1} and \cite{ka2}). It can be recalled that a non-negative function $f:\mathcal{R}^2\longmapsto\mathcal{R}_+$ is said to be $TP_2$ if
\begin{equation}\label{eq1a}
	f(u_1,v_1)f(u_2,v_2)-f(u_2,v_1)f(u_1,v_2)\geq 0,
\end{equation}
for all $u_1\leq u_2$ and $v_1\leq v_2$. This $f$ is said to be $RR_2$ if the inequality in (\ref{eq1a}) is reversed. This can equivalently be written as follows\\
$f$ is $TP_2$ ($RR_2$) if $f(u,v_2)/f(u,v_1)$ is increasing (decreasing) in $u$ for all $v_1\leq v_2$. \\
The four Propositions below are used to prove the main results. The proofs can be found in Kundu \emph{et al.}~\cite{ku}.
\begin{p1}\label{le1} For any positive integer $n$ and positive real
	 number $x$, let $K_n(x)>0$ be a $RR_2$ function in $n\in \mathcal{N}^+$ and $x\in \mathcal{R}_+$. Assume that any function $f_n(x)$ be such that 
\begin{itemize}
\item [i)] for each $x\in \mathcal{R}_+$, $f_n(x)$ changes sign at most once and if the change occurs, it is from negative to positive as $n$ traverses in $\mathcal{N}^+$;
\item [ii)] for each $n\in \mathcal{N}^+$, $f_n(x)$ decreases in $x$;
\item [iii)] $w(x)=\sum_{n=1}^{\infty}{f_n(x)K_n(x)}$ converges absolutely and defines a continuous function of $x$.
\end{itemize}
Then $w(x)$ changes sign at most once and if the change occurs, it is from positive to negative.
\end{p1} 

\begin{p1} \label{le2}
For any positive integer $n$ and positive real number $x$, let $K_n(x)>0$ be a $RR_2$ function in $n\in \mathcal{N}^+$ and $x\in \mathcal{R}_+$. Assume that any function $f_n(x)$ be such that 
\begin{itemize}
\item [i)] for each $x\in \mathcal{R}_+$, $f_n(x)$ changes sign at most once and if the change occurs, it is from positive to negative as $n$ traverses in $\mathcal{N}^+$;
\item [ii)] for each $n\in \mathcal{N}^+$, $f_n(x)$ increases in $x$;
\item [iii)] $w(x)=\sum_{n=1}^{\infty}{f_n(x)K_n(x)}$ converges absolutely and defines a continuous function of $x$.
\end{itemize}
Then $w(x)$ changes sign at most once and if the change occurs, it is from negative to positive.
\end{p1} 

\begin{p1}\label{le3}
For any positive integer $n$ and positive real number $x$, let $K_n(x)>0$ be a $TP_2$ function in $n\in \mathcal{N}^+$ and $x\in \mathcal{R}_+$. Assume that any function $f_n(x)$ be such that 
\begin{itemize}
\item [i)] for each $x\in \mathcal{R}_+$, $f_n(x)$ changes sign at most once and if the change occurs, it is from positive to negative as $n$ traverses in $\mathcal{N}^+$;
\item [ii)] for each $n\in \mathcal{N}^+$, $f_n(x)$ decreases in $x$;
\item [iii)] $w(x)=\sum_{n=1}^{\infty}{f_n(x)K_n(x)}$ converges absolutely and defines a continuous function of $x$.
\end{itemize}
Then $w(x)$ changes sign at most once and if the change occurs, it is from positive to negative.
\end{p1} 

\begin{p1}\label{le4}
For any positive integer $n$ and positive real number $x$, let $K_n(x)>0$ be a $TP_2$ function in $n\in \mathcal{N}^+$ and $x\in \mathcal{R}_+$. Assume that any function $f_n(x)$ be such that 
\begin{itemize}
\item [i)] for each $x\in \mathcal{R}_+$, $f_n(x)$ changes sign at most once and if the change occurs, it is from negative to positive as $n$ traverses in $\mathcal{N}^+$;
\item [ii)] for each $n\in \mathcal{N}^+$, $f_n(x)$ increases in $x$;
\item [iii)] $w(x)=\sum_{n=1}^{\infty}{f_n(x)K_n(x)}$ converges absolutely and defines a continuous function of $x$.
\end{itemize}
Then $w(x)$ changes sign at most once and if the change occurs, it is from negative to positive.
\end{p1} 

Kundu \emph{et al.}~\cite{ku} also provide counterexamples to show that, for the combinations of conditions listed below, no definitive conclusion can be made regarding the sign change behavior of $w(x).$
\begin{itemize} 
\item[\textbf{Case-I}]   $K_n(x)$  is  $RR_2$ in $n\in \mathcal{N}^+$ and $x \in \mathcal{R}_+$; $f_n(x)$ is increasing in  $x \in \mathcal{R}_+$ for all $n \in \mathcal{N}^+$;  and  $f_n(x)$ has at most one change of sign from negative to positive as $n$ traverses $\mathcal{N}^+.$
\item[\textbf{Case-II}]   $K_n(x)$  is  $RR_2$ in $n\in \mathcal{N}^+$ and $x \in \mathcal{R}_+$;  $f_n(x)$ is decreasing in  $x \in \mathcal{R}_+$  for all $n \in \mathcal{N}^+$; and  $f_n(x)$ has at most one change of sign from  positive to negative as $n$ traverses $\mathcal{N}^+.$
\item[\textbf{Case-III}]   $K_n(x)$  is  $TP_2$ function in $n\in \mathcal{N}^+$ and $x \in \mathcal{R}_+$;   $f_n(x)$ is decreasing in  $x \in \mathcal{R}_+$ for all $n \in \mathcal{N}^+$; and  $f_n(x)$ has at most one change of sign from  negative to positive as $n$ traverses $\mathcal{N}^+.$
\item[\textbf{Case-IV}]   $K_n(x)$  is  $TP_2$ function in $n\in \mathcal{N}^+$ and $x \in \mathcal{R}_+$;  $f_n(x)$ is increasing in  $x \in \mathcal{R}_+$  for all $n \in \mathcal{N}^+$; and  $f_n(x)$ has at most one change of sign   positive to negative  as $n$ traverses $\mathcal{N}^+.$
\end{itemize}

\section{Main Results}
\setcounter{equation}{0}

Let $X_i$ ($Y_i$), $i=1,2,\ldots,n$, be $n$ number of independent random variables having distribution functions $F_i\left(x\right)$ ($G_i\left(x\right)$), for all $i=1,2,\ldots,n$ having supports $(l_x^i,u_x^i)((l_y^i,u_y^i))$. Also let $\bar F_{n:n}\left(x\right)$ and $\bar G_{n:n}\left(x\right)$ ($ F_{1:n}\left(x\right)$ and $ G_{1:n}\left(x\right)$) be the survival ( distribution) functions of $X_{n:n}$ and $Y_{n:n}$ ($X_{1:n}$ and $Y_{1:n}$) respectively with $X_{1:n}$ and $X_{n:n}$ being defined as $$ X_{1:n}=\min\left\{X_1, X_2,...,X_n\right\}$$ and $$ X_{n:n}=\max\left\{X_1, X_2,...,X_n\right\}.$$ Also let, for $i=1,2,\cdots,n$, $u=\max\{u_x^i,u_y^i \}$ and $l=\min\{l_x^i,l_y^i \}$. Let $N$ be a discrete random variable with probability mass function (p.m.f) $p(n)$, $n\in\mathcal{N}^+$. Now, for the random variable $N$, if $\bar F_{N:N}\left(x\right)$ and $\bar G_{N:N}\left(x\right)$ ($ F_{1:N}\left(x\right)$ and $G_{1:N}\left(x\right)$) denote the survival (distribution) functions of $X_{N:N}$ and $Y_{N:N}$ ($X_{1:N}$ and $Y_{1:N}$) respectively, then it can be written that 
\begin{equation}\label{eq2}
	\bar F_{N:N}(x)=\sum_{n=1}^\infty \bar F_{n:n}(x)p(n),\ \bar G_{N:N}(x)=\sum_{n=1}^\infty \bar G_{n:n}(x)p(n),
\end{equation}
\begin{equation}\label{eq3}
	 F_{1:N}(x)=\sum_{n=1}^\infty F_{1:n}(x)p(n)\ \text{and}\ G_{1:N}(x)=\sum_{n=1}^\infty  G_{1:n}(x)p(n).
\end{equation}

The theorems below derive a set of sufficient conditions which ensure preservation of $rh$ and $hr$ ordering between minimum and maximum order statistics, respectively.

\begin{t1}\label{th1}
	Suppose $X_{1:n}\sim F_{1:n}(x)$ and  $Y_{1:n}\sim G_{1:n}(x).$ Also let the support of a positive integer valued random variable $N$ having pmf $p(n)$ be $\mathcal{N}^+$. Now, for all $x\geq l$, if $\frac{{F}_{1:n}(x)}{{G}_{1:n}(x)}$ is increasing in $n\in\mathcal{N}^+$, then $X_{1:n}\leq_{rh} Y_{1:n}$ implies $X_{1:N}\leq_{rh} Y_{1:N}.$
	\end{t1}
	{\bf Proof:} Here, we need to prove that $\frac{ {F}_{1:N}(x)}{{G}_{1:N}(x)}$ is decreasing in $x$ or equivalently $\frac{\sum_{n=1}^{\infty}{{F}_{1:n}(x)p(n)}}{\sum_{n=1}^{\infty}{{G}_{1:n}(x)p(n)}}$ is decreasing in $x.$ Now for any $\lambda>0$, let 
	\begin{eqnarray*}
w_3(x) &=&\sum_{n=1}^{\infty}{\left[{F}_{1:n}(x)-\lambda{G}_{1:n}(x)\right]p(n)}\\
&=& \sum_{n=1}^{\infty}{{G}_{1:n}(x)\left[\frac{{F}_{1:n}(x)}{{G}_{1:n}(x)}-\lambda\right]p(n)}\\
&=& \sum_{n=1}^{\infty}{{G}_{1:n}(x) f_n(x) p(n)} \ \text{(say),}
\end{eqnarray*}
where $f_n(x)=\left[\frac{{F}_{1:n}(x)}{{G}_{1:n}(x)}-\lambda\right]$.\\
Now, $X_{1:n}\leq_{rh} Y_{1:n}$ implies that $\frac{{F}_{1:n}(x)}{{G}_{1:n}(x)}$ is decreasing in $x$, and hence $f_n(x)$ is also decreasing in $x.$ \\
Since $\frac{{F}_{1:n}(x)}{{G}_{1:n}(x)}$ is increasing in $n$, it is clear that $f_n(x)$ has at most one change of sign as $n$ traverses and if it does occur, it is from negative to positive.\\ 
Also, Theorem \ref{theo11} concludes that for any two positive integers $n_1\leq n_2$, $Y_{1:n_1}\geq_{rh}Y_{1:n_2}$, which by definition of $rh$ ordering indicates that $\frac{{G}_{1:n_1}(x)}{{G}_{1:n_2}(x)}$ is increasing in $x$, yielding that ${G}_{1:n}(x)$ is $RR_2$ function in $n$ and $x$.\\
Thus, by Proposition $\ref{le1}$, $w_3(x)$ changes its sign at most once and if the changes occurs, it is from positive to negative. Now, as we know that, for any $x\geq 0$, $f(x)-cg(x)$ has at most one change of sign from  positive to negative can equivalently  written as $\frac{f(x)}{g(x)}$ is decreasing in $x\in (0,\infty)$, thus it can be concluded that $\frac{\sum_{n=1}^{\infty}{{F}_{1:n}(x)}p(n)}{\sum_{n=1}^{\infty}{G}_{1:n}(x)p(n)}$ is decreasing in $x$, proving the result.$\hfill\Box$\\\\
Next, we illustrate Theorem \ref{th1} with an example.\\ 
\begin{ex1}\label{exam1}
Let, $X\sim\bar F(x)$ be a random variable having failure rate function $r(x)$. Also let, for $n\in\mathcal{N}^+$, $X_i\sim \bar F^{\lambda_i}\left(x\right)$ and $Y_i\sim \bar F^{\mu_i}\left(x\right),~i=1,2,...,n$ be two sequences of $inid$ random variables where $\lambda_i\geq\mu_i$. Then, the survival and reversed hazard rate functions of $X_{1:n}$ and $Y_{1:n}$ are obtained as  
\begin{equation}\label{eq4}
	\bar F_{1:n}(x)=\prod_{i=1}^n\bar F^{\lambda_i}\left(x\right)=\bar F^{\lambda}\left(x\right),\ \bar G_{1:n}(x)=\bar F^{\mu}\left(x\right),
\end{equation}
and 
\begin{equation}\label{eq5}
	\bar r_{1:n}(x)=\frac{\lambda r(x) }{\bar F^{-\lambda}\left(x\right)-1}\ \text{and}\ \bar s_{1:n}(x)=\frac{\mu r(x) }{\bar F^{-\mu}\left(x\right)-1}
\end{equation}
respectively, where $\lambda=\sum_{i=1}^n\lambda_i$ and $\mu=\sum_{i=1}^n\mu_i$. \\
Since, $\lambda_i\geq\mu_i$ and thus $\lambda\geq\mu$, using Lemma \ref{lemma12} it can be written  from (\ref{eq5}) that $\bar r_{1:n}(x)\leq\bar s_{1:n}(x),$
proving that $X_{1:n}\leq_{rh}Y_{1:n}$.\\
Now, let us assume that $X$ follows exponential distribution with survival function $\bar F(x)=e^{-x}, x\geq 0.$ Also let, for $i=1,2\ldots, 5$  $X_i\sim \bar F^{\lambda_i}\left(x\right)$ and $Y_i\sim \bar F^{\mu_i}\left(x\right)$ where\\ $\left(\lambda_1,\lambda_2,\lambda_3,\lambda_4,\lambda_5\right)=(0.1,0.2,0.25,0.35,0.5)$ and $\left(\mu_1,\mu_2,\mu_3,\mu_4,\mu_5\right)=(0.05,0.15,0.23,0.33,0.5)$, which implies that $\lambda_i\geq\mu_i$. Now, if it is assumed that  $$X_1, X_2, X_3~(Y_1, Y_2, Y_3)~\text{are selected with probability } P(N=3)=p(3)=1/5,$$ $$X_1, X_2, X_3, X_4 ~(Y_1, Y_2, Y_3,Y_4)~\text{are selected with probability } P(N=4)=p(4)=2/5~\text{and}$$ $$X_1, X_2, X_3, X_4, X_5~(Y_1, Y_2, Y_3, Y_4, Y_5)~\text{are selected with probability } P(N=5)=p(5)=2/5,$$
then from Figure \ref{figure1} it can be concluded that $\frac{F_{1:n}(x)}{G_{1:n}(x)}$ is increasing in $n$, and Figure \ref{figure11} shows that $\frac{F_{1:N}(x)}{G_{1:N}(x)}$ is decreasing in $x$, proving that $X_{1:N}\leq_{rh}Y_{1:N}$. Here substitution  $x=-\ln y$ is used to capture whole real line.
\begin{figure}[ht]
	\centering
	\includegraphics[height=6 cm]{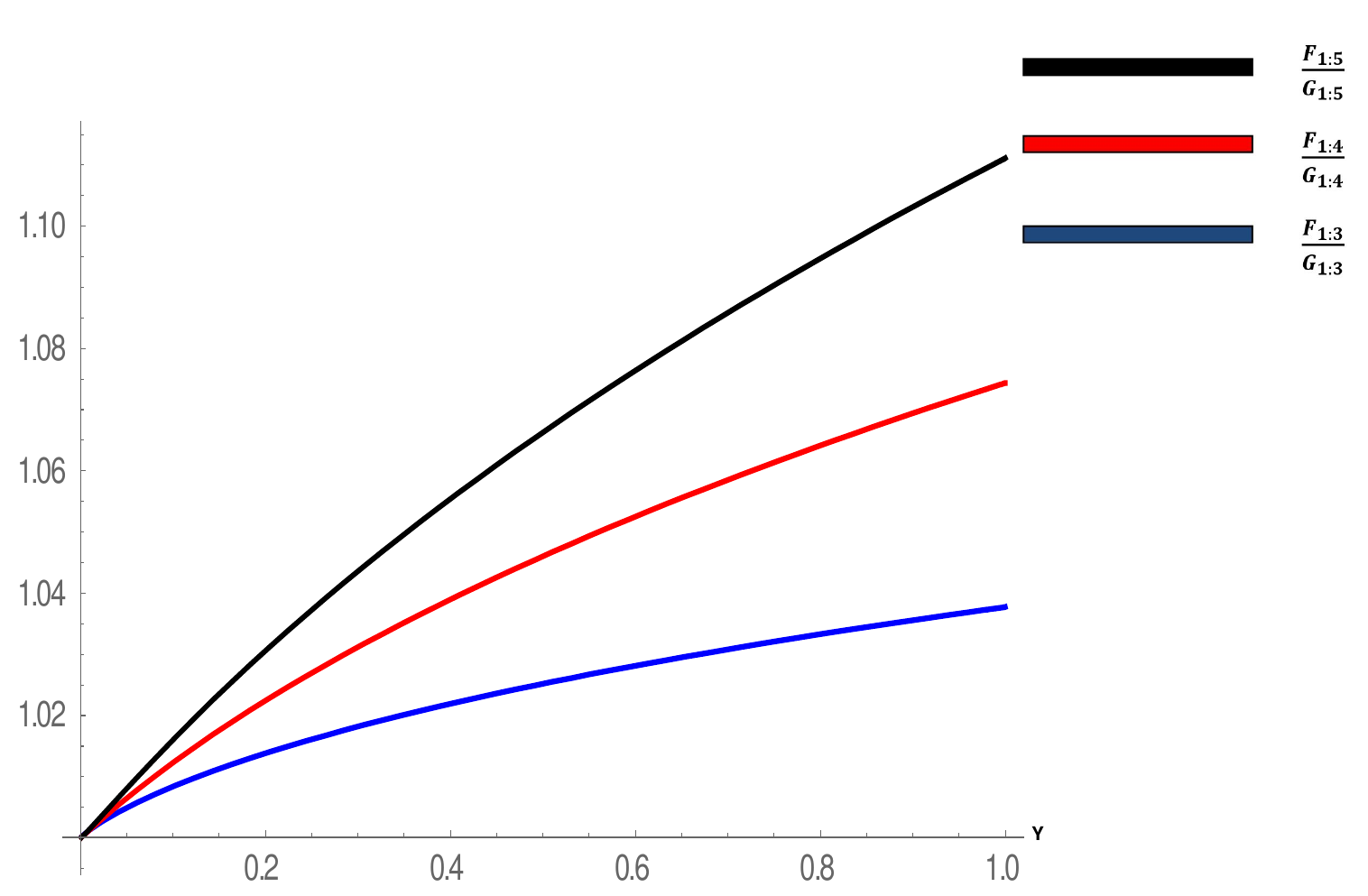}
	\caption{Graph of $\frac{{F}_{1:n}(-\ln y)}{{G}_{1:n}(-\ln y)}$, Example \ref{exam1}} \label{figure1}
\end{figure}
\begin{figure}[ht]
	\centering
	\includegraphics[height=6 cm]{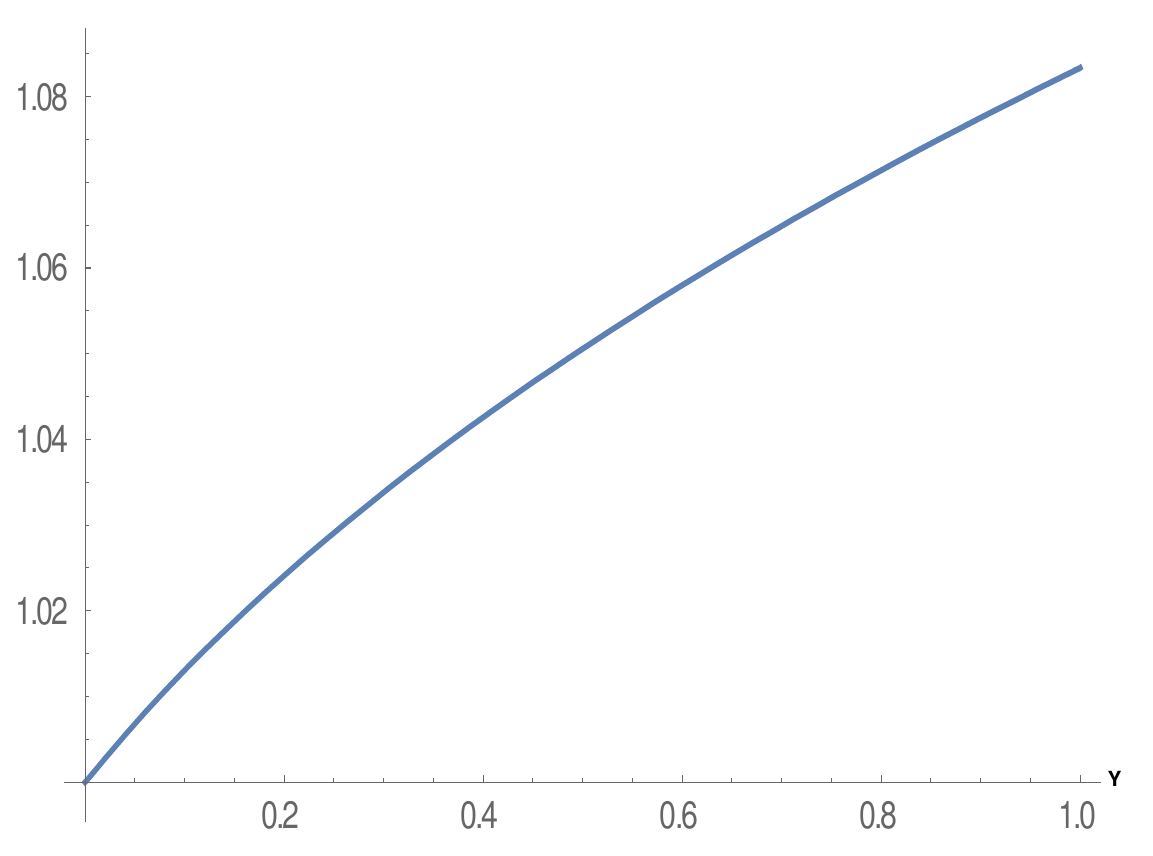}
	\caption{Graph of $\frac{{F}_{1:N}(-\ln y)}{{G}_{1:N}(-\ln y)}$, Example \ref{exam1}} \label{figure11}
\end{figure}

\end{ex1}

The following conclusion can be drawn from Theorem \ref{th1} on the $iid$ case.
\begin{c1}\label{cor}
Let us consider \( X_i \sim \bar{F}_i(x) = \bar{F}(x) \) and \( Y_i \sim \bar{G}_i(x) = \bar{G}(x) \) for \( i = 1, 2, \ldots, n \). It can be easily concluded that for all $x\geq0,$ if $\frac{F_{1:n}(x)}{G_{1:n}(x)} = \left( \frac{F(x)}{G(x)} \right)^n$ is increasing in $n\in\mathcal{N}^+$, then $X_{1:n}\leq_{rh} Y_{1:n}$ implies $X_{1:N}\leq_{rh} Y_{1:N}.$ Hence, $rh$ ordering between minimum order statistics is preserved.
\end{c1}
	 
\begin{r1}\label{rem}
 For $i=1,2,\ldots,n$, if $X_i\sim  \bar F_i(x)=\bar F(x)$ and  $Y_i\sim  \bar G_i(x)=\bar G(x)$ are two sets of $n$ number of $iid$ random variables, then the condition in Theorem $3.1$ of Kundu \emph{et al.}~\cite{ku} that $\frac{\bar{F}_{1:n}(x)}{\bar{G}_{1:n}(x)}$ is increasing in $n$ contradicts the fact that \( X_{1:n} \leq_{hr} Y_{1:n} \) for all \( n \). In fact, since for all n, \( X_{1:n} \leq_{hr} Y_{1:n} \) implies $\bar{F}(x)\leq\bar{G}(x)$ for all $x\geq 0$, it follows that $\frac{\bar{F}_{1:n}(x)}{\bar{G}_{1:n}(x)}$ must be decreasing in $n.$ Therefore, by Case II of Counterexample~\ref{le1} in Kundu \emph{et al.}~\cite{ku}, it can be concluded that, no definitive conclusion can be drawn regarding the preservation of $hr$ ordering between $X_{1:N}$ and $Y_{1:N}.$\\
\end{r1}
	
Using similar reasoning as in Corollary \ref{cor}, it can be shown that the following theorems are applicable to their respective $iid$ cases. However, based on the arguments presented in Remark \ref{rem}, Theorems \ref{th2}~-~\ref{th4} of Kundu \emph{et al.}~\cite{ku}, which pertain to the non-identical cases, do not apply to the identical cases.

\begin{t1}\label{th2}
	Suppose $X_{1:n}\sim  F_{1:n}(x)$ and  $Y_{1:n}\sim G_{1:n}(x).$ Also let the support of a positive integer valued random variable $N$ having pmf $p(n)$ be $\mathcal{N}^+$. Now, for all $ x\geq l$, if $\frac{{F}_{1:n}(x)}{{G}_{1:n}(x)}$ is decreasing in $n\in\mathcal{N}^+$, then $X_{1:n}\geq_{rh} Y_{1:n}$ implies $X_{1:N}\geq_{rh} Y_{1:N}.$
\end{t1}	
{\bf Proof}: The theorem can be proved using Proposition \ref{le2}, Theorem \ref{theo11} and following the proof of Theorem \ref{th1}.\\

 \begin{t1}\label{th3}
 	Suppose $X_{n:n}\sim \bar F_{n:n}(x)$ and  $Y_{n:n}\sim \bar G_{n:n}(x).$  Also let the support of a positive integer valued random variable $N$ having pmf $p(n)$ be $\mathcal{N}^+$. Now, for all $x\leq u$, if $r_{Y_{n:n}}(x)$, the hazard rate function of $Y_{n:n}$ is decreasing in $n\in\mathcal{N}^+$ and $\frac{\bar F_{n:n}(x)}{\bar G_{n:n}(x)}$ is decreasing in $n\in\mathcal{N}^+$, then $X_{n:n}\leq_{hr} Y_{n:n}$ implies $X_{N:N}\leq_{hr} Y_{N:N}.$
 \end{t1}	
 {\bf Proof}: If $s_{n:n}(x)$ is decreasing in $n$, then for any two positive integers $n_1\leq n_2$ and for all $x\leq u$ it can be written that $r_{Y_{n_1:n_1}}(x)\geq r_{Y_{n_2:n_2}}(x)$, which equivalently concluded that $\frac{\bar G_{n_1:n_1}(x)}{\bar G_{n_2:n_2}(x)}$ is decreasing in $x\leq u$. Thus, the theorem can be proved using Proposition~$\ref{le3}$ and following the line of the proof of Theorem \ref{th4}.\\
 
\begin{t1}\label{th4}
 	Suppose $X_{n:n}\sim \bar F_{n:n}(x)$ and  $Y_{n:n}\sim \bar G_{n:n}(x).$ Also let the support of a nonnegative integer valued random variable $N$ having pmf $p(n)$ be $\mathcal{N}^+$. Now, for all $x\leq u$, if $r_{Y_{n:n}}(x)$, the hazard rate function of $Y_{n:n}$ is decreasing in $n\in\mathcal{N}^+$ and $\frac{\bar F_{n:n}(x)}{\bar G_{n:n}(x)}$ is increasing in $n\in\mathcal{N}^+$, then $X_{n:n}\geq_{hr} Y_{n:n}$ implies $X_{N:N}\geq_{hr} Y_{N:N}.$
 	 \end{t1}
 {\bf Proof}: The theorem can be proved using Proposition~\ref{le4} and following the proof of Theorem \ref{th3}.\\
 	
\begin{figure}[ht]
 	\centering
 	\includegraphics[height=5 cm, width=10 cm]{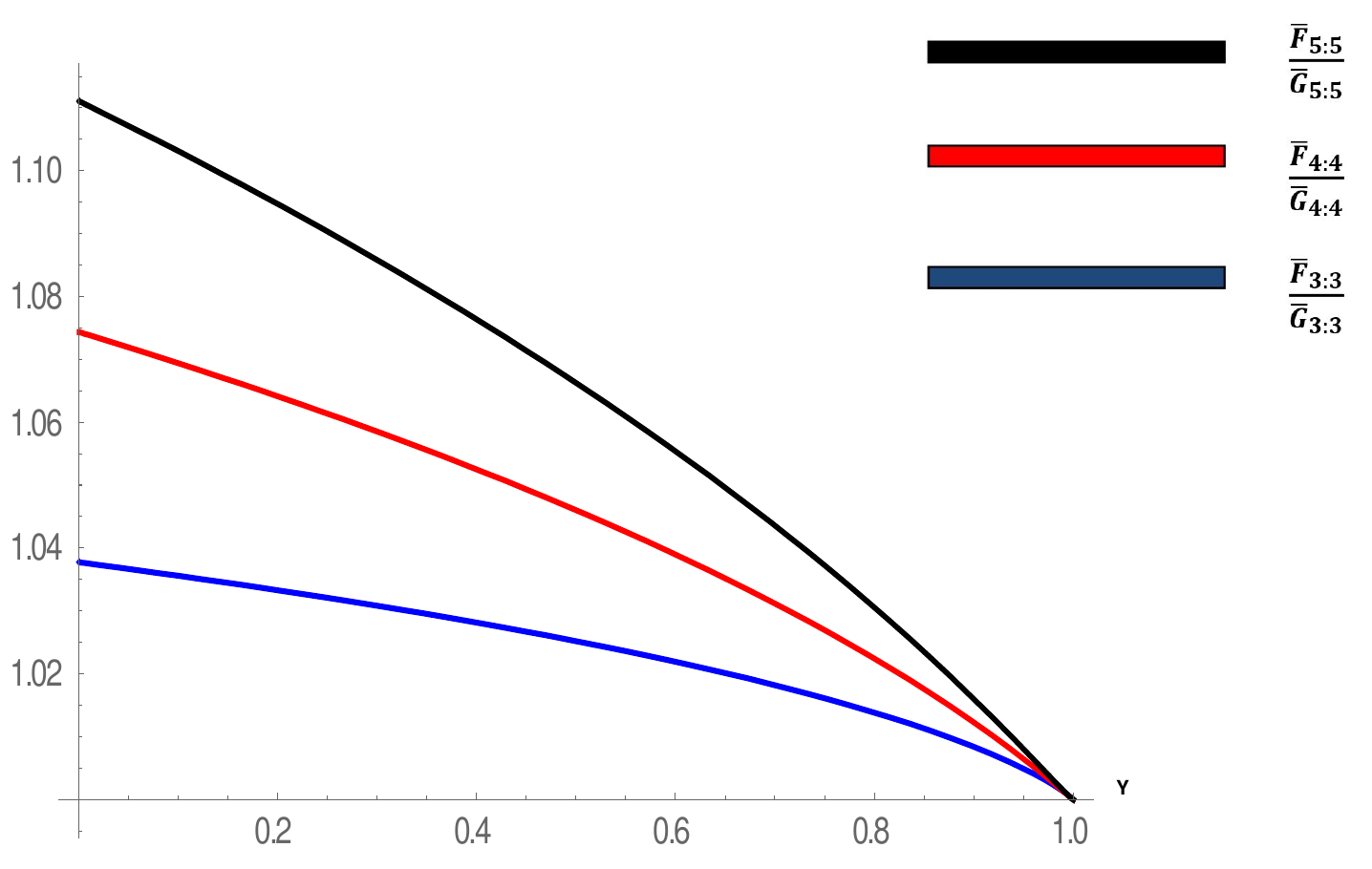}
 	\caption{Graphs of $\frac{\bar F_{n:n}(-\ln y)}{\bar G_{n:n}(-\ln y)}$, Example \ref{exam2}} \label{figure2.1a}
 \end{figure}

\begin{figure}[ht]
 	\centering
 	\includegraphics[height=5 cm,, width=10 cm]{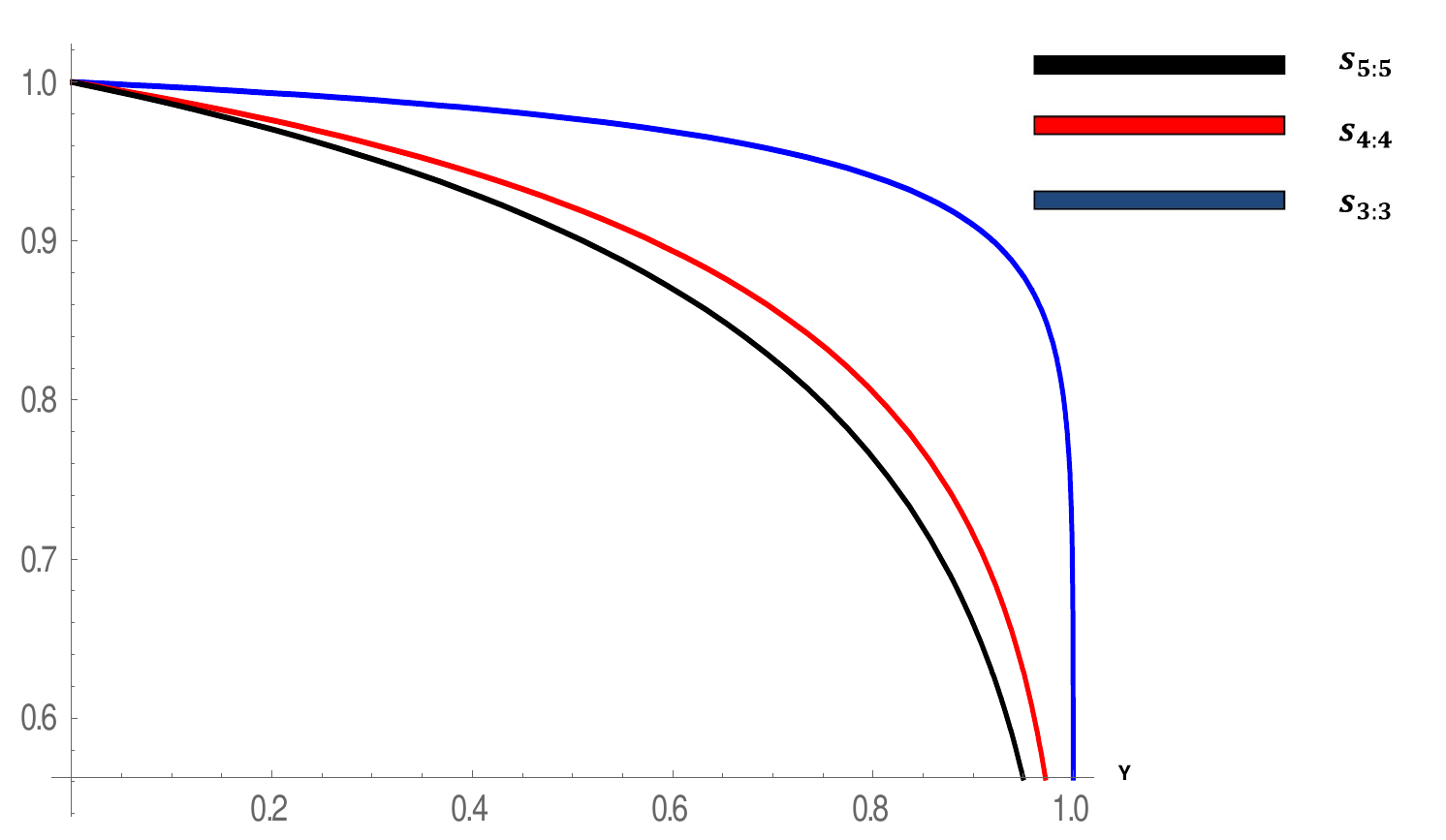}
 	\caption{Graphs of  $s_{n:n}(-\ln y)$, Example \ref{exam2}} \label{figure2.1b}
 \end{figure}

\begin{figure}[ht]
 	\centering
 	\includegraphics[height=5 cm]{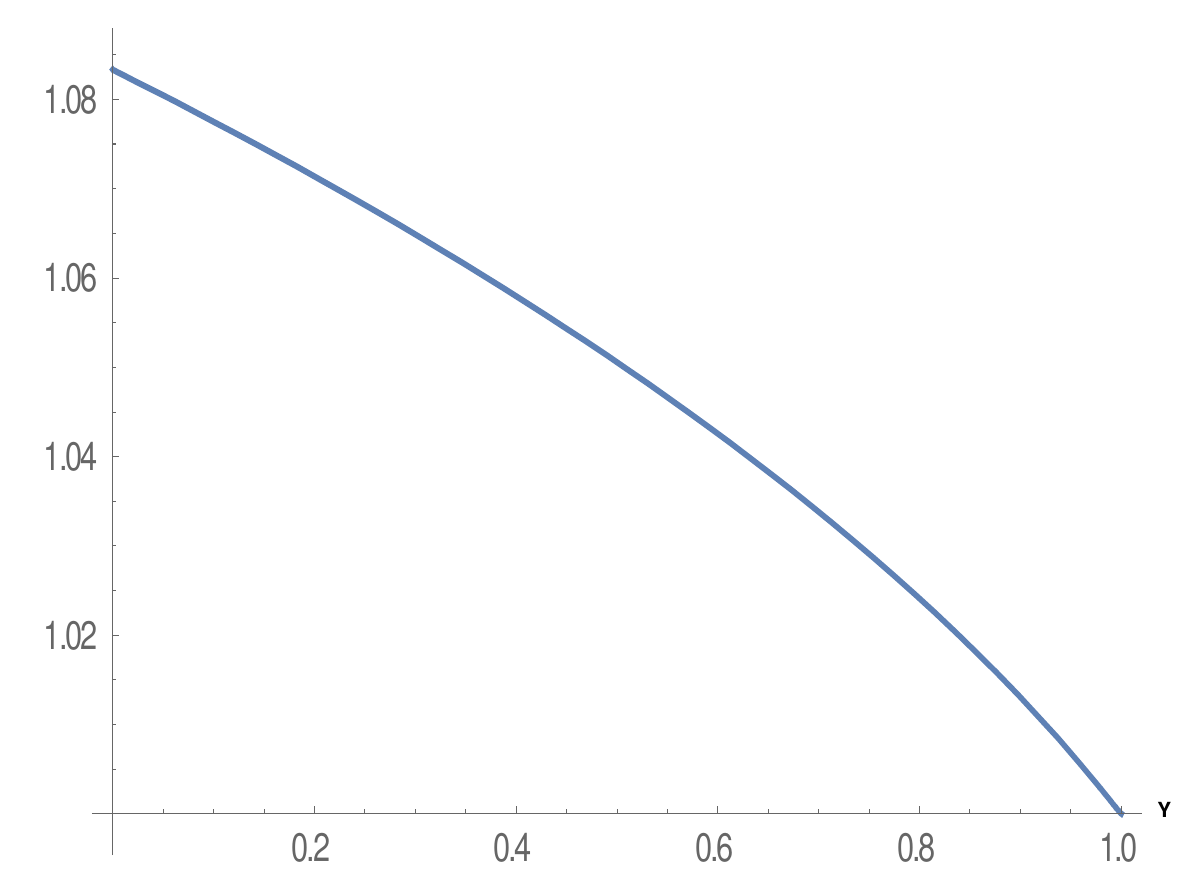}
 	\caption{Graph of \( \frac{\bar F_{N:N}(-\ln y)}{\bar G_{N:N}(-\ln y)} \), Example \ref{exam2}} \label{figure2.1c}
 \end{figure}

The example, given below, illustrates Theorem~\ref{th4}.
 	 \begin{ex1}\label{exam2}
 	Let, $X\sim F(x)$ be a random variable having reversed hazard rate function $\bar r(x)$. Also let, for $n\in\mathcal{N}^+$, $X_i\sim F^{\lambda_i}\left(x\right)$ and $Y_i\sim F^{\mu_i}\left(x\right),~i=1,2,...,n$ be two sequences of $inid$ random variables where $\lambda_i\geq\mu_i$. Then, the distribution and hazard rate functions of $X_{n:n}$ and $Y_{n:n}$ are obtained as  
 		\begin{equation}\label{eq4a}
 			F_{n:n}(x)=\prod_{i=1}^n F^{\lambda_i}\left(x\right)= F^{\lambda}\left(x\right),\ G_{n:n}(x)=F^{\mu}\left(x\right),
 		\end{equation}
 		and 
 		\begin{equation}\label{eq5a}
 			r_{n:n}(x)=\frac{\lambda \bar r(x) }{F^{-\lambda}\left(x\right)-1}\ \text{and}\ s_{n:n}(x)=\frac{\mu \bar r(x) }{F^{-\mu}\left(x\right)-1}
 		\end{equation}
 		respectively, where $\lambda=\sum_{i=1}^n\lambda_i$ and $\mu=\sum_{i=1}^n\mu_i$. \\
 		Since, $\lambda_i\geq\mu_i$ and thus $\lambda\geq\mu$, using Lemma \ref{lemma12} it can be written  from (\ref{eq5a}) that $r_{n:n}(x)\leq s_{n:n}(x),$
 		proving that $X_{n:n}\geq_{hr}Y_{n:n}$.\\
 		Now, let us assume that $X$ follows exponential distribution with distribution function $F(x)=1-e^{-x}, x\geq 0.$ Also let, for $i=1,2\ldots, 5$, suppose $X_i\sim F^{\lambda_i}\left(x\right)$ and $Y_i\sim F^{\mu_i}\left(x\right)$ where each $\lambda_i$ and $\mu_i$	are same as of Example~\ref{exam1}. Now, taking the same values of $p(n)$ for the same selections of $X_is$ ($Y_is$) as that of Example \ref{exam1}, it can be seen from Figure~\ref{figure2.1a} that $\frac{\bar F_{n:n}(x)}{\bar G_{n:n(x)}}$ is increasing in $n$ and $x$. Also Figure~\ref{figure2.1b} ensures that $s_{n:n}(x)$ is decreasing in $n$. So all conditions of Theorem~\ref{th4} are satisfied. Figure~\ref{figure2.1c} shows that $\frac{\bar F_{N:N}(x)}{\bar G_{N:N}(x)}$ is increasing in $x$, proving the result of Theorem~\ref{th4}. Here also substitution $x=-\ln y$ is used to capture whole real line.
 	
 \end{ex1}
 
\begin{r1}\label{rem1}
In Theorem $3.5$ of Kundu \emph{et al.}~\cite{ku}, it is shown that for $inid$ random variables, the $lr$ oredring between minimum order statistics is preserved under a set of sufficient conditions. Specifically, the theorem states that ``if \( \frac{g_{1:n}(x)}{f_{1:n}(x)} \) is increasing in \( n \) for all \( l \leq x \leq u \), and for \( n_1 \leq n_2 \), \( X_{1:n_1} \geq_{lr} X_{1:n_2} \), then \( X_{1:n} \geq_{lr} Y_{1:n} \) implies \( X_{1:N} \geq_{lr} Y_{1:N} \)." However, for $iid$ random variables, this theorem cannot be universally applied. The reason is as follows:\\

Consider \( X_i \sim \bar{F}(x) \) and \( Y_i \sim \bar{G}(x)\) for \( i = 1, 2, \ldots, n \). The density of $X_{1:n}$ and $Y_{1:n}$ are given by\\ 
 \[
 f_{1:n}(x) = n \bar{F}^{n-1}(x) f(x) \quad \text{and} \quad g_{1:n}(x) = n \bar{G}^{n-1}(x) g(x),
 \]
 which leads to\\
 \[
 \frac{g_{1:n}(x)}{f_{1:n}(x)} = \left( \frac{\bar{G}(x)}{\bar{F}(x)} \right)^{n-1} \frac{g(x)}{f(x)}.
 \]
Here, it can be observed that the condition of Theorem $3.5$ in Kundu \emph{et al.}~\cite{ku}, which states that $\frac{g_{1:n}(x)}{f_{1:n}(x)} \text{ is increasing in } n\in\mathcal{N}^+,$ will hold if and only if for all $x\geq 0,$ $\bar{F}(x)\leq\bar{G}(x)$. This can be equivalently written as $$\left[\bar{F}(x)\right]^n\leq\left[\bar{G}(x)\right]^n for all n\in\mathcal{N}^+,$$ which implies \( X_{1:n} \leq_{st} Y_{1:n} \). This fact contradicts another assumption of the theorem, which states that \( X_{1:n} \geq_{lr} Y_{1:n} \). Therefore, for $iid$ random variables, the preservation of $lr$ ordering  between minimum order statistics cannot be guaranteed by Theorem $3.5$ of Kundu \emph{et al.}~\cite{ku}\\ 

\hspace*{0.2 in} To address this issue, an alternative version of this theorem could be stated as follows:  \\
``if \( \frac{g_{1:n}(x)}{f_{1:n}(x)} \) is decreasing in \( n \), and for all \( n_1 \leq n_2 \), \( X_{1:n_1} \leq_{lr} X_{1:n_2} \), then \( X_{1:n} \geq_{lr} Y_{1:n} \) implies \( X_{1:N} \geq_{lr} Y_{1:N} \).''\\
However, upon further examination, it can be seen that the condition \( X_{1:n_1} \leq_{lr} X_{1:n_2} \) for \( n_1 \leq n_2 \) does not hold good. Thus, it can be concluded that for $iid$ random variables, no general conclusion can be drawn about the preservation of $lr$ ordering between \( X_{1:N} \) and \( Y_{1:N} \). \\

\hspace*{0.2 in} Following similar reasoning, for Theorems $3.6$~-~$3.8$ of Kundu \emph{et al.}~\cite{ku}, it can also be concluded that for $iid$ random variables, no conclusion can be drawn regarding the preservation of $lr$ ordering between the maximum order statistics.
 \end{r1}

\section{Conclusion}
\setcounter{equation}{0}
This short article revisits the work of Kundu \emph{et al.,}~\cite{ku} which presents results on the comparison of random extremes for $inid$ random variables. The contribution of this short communication are as follows:  
\begin{itemize}
\item Kundu \emph{et al.}~\cite{ku} provided results on $hr$ and $rh$ ordering for random minimum and maximum order statistics, respectively. However, these results are shown to be inapplicable to the respective $iid$ cases. 
\item This article presents new results on $hr$ and $rh$ ordering for random maximum and minimum order statistics. 
\item It is shown that the new results are applicable to the respective $iid$ cases. 
 \item Kundu \emph{et al.}~\cite{ku} provided results on $lr$ ordering for random minimum and maximum order statistics. It is found that no sufficient conditions exist for the preservation of $lr$ ordering between the random extremes in the $iid$ case.  
\end{itemize}

\section*{Acknowledgment}
Research scholarship grant from National Board for Higher Mathematics(NBHM)(Ref No. 0203/11/2019-R$\&$D-II/9253) is acknowledgment by Bidhan Modok.

\end{document}